\documentclass[twocolumn,superscriptaddress]{revtex4-2}
\usepackage{graphicx}
\usepackage{tikz}
\usepackage{amssymb,amsmath}
\usepackage{bm}
\usepackage{dcolumn}
\usepackage{mathtools}

\usepackage{hyperref}
\hypersetup{colorlinks=true,breaklinks,urlcolor=blue,linkcolor=blue,citecolor=blue}
\usepackage{color}

\usepackage[capitalise]{cleveref}
\usepackage[notrig]{physics}   
\usepackage{braket}

\usepackage{soul}

\newcommand{\rom}[1]{\uppercase\expandafter{\romannumeral #1}}
\renewcommand{\tr}{\mathrm{Tr}}

\renewcommand{\hat}[1]{#1}

\makeatletter
\def\maketitle{
	\@author@finish
	\title@column\titleblock@produce
	\suppressfloats[t]}
\makeatother

\begin{document}

\title{Canonically consistent quantum master equation}
\author{Tobias Becker}
\email{tobias.becker@tu-berlin.de}
\author{Alexander Schnell}
\email{schnell@tu-berlin.de}
\affiliation{Institut f\"ur Theoretische Physik, Technische Universit\"at Berlin, Hardenbergstr.~36, D-10623 Berlin, Germany}
\author{Juzar Thingna}
\email{juzar\_thingna@uml.edu}
\thanks{\\AS and JT contributed equally to this work}
\affiliation{Department of Physics and Applied Physics, University of Massachusetts, Lowell, MA 01854, USA}
\affiliation{Center for Theoretical Physics of Complex Systems, Institute for Basic Science (IBS), Daejeon 34126, Republic of Korea}

\date{\today}

\begin{abstract}
We put forth a new class of quantum master equations that correctly reproduce the asymptotic state of an open quantum system beyond the infinitesimally weak system-bath coupling limit. Our method is based on incorporating the knowledge of the reduced steady state into its dynamics. The correction not only steers the reduced system towards a correct  steady state but also improves the accuracy of the dynamics, thereby refining the archetypal Born-Markov  weak-coupling second-order master equations. In case of equilibrium, since a closed form for the steady state exists in terms of a mean force Gibbs state, we utilize this form to correct the Redfield quantum master equation. Using an exactly solvable harmonic oscillator system we benchmark our approach with the exact solution showing that our method also helps correcting the long-standing issue of positivity violation, albeit without complete positivity. Our method of a canonically consistent quantum master equation, opens a new perspective in the theory of open quantum systems leading to a reduced density matrix accurate beyond the commonly used Redfield and Lindblad equations, while retaining the same conceptual and numerical complexity.
\end{abstract}

\maketitle

\noindent \textit{Introduction.--}
The central problem of the theory of open quantum systems is to describe the dynamics of a quantum system in contact with a reservoir~\cite{breuerpetruccione, Weiss, carmichael}. Most applications describe the dynamics using a weak-coupling quantum master equation (QME) such as Lindblad~\cite{breuerpetruccione,GLindblad1976, GoriniKossakowski1976} or Redfield~\cite{carmichael,breuerpetruccione,AGRedfield65} that are valid under a stringent set of assumptions. Going beyond this standard weak-coupling approach is tedious and despite formal equations available since the past 45 years~\cite{Nakajima58, Zwanzig60, FULINSKI1968, Shibata77}, there exist only a handful of model-independent practical methods that go beyond weak coupling~\cite{HEOM89,BLaird1991,QUAPI95,CaoJCP,Timm2011,JThingnaWJSheng2014,RCNazir14,TTM14,SAlipour2019,RN1,Imamoglu94,Garraway97,Mazzola09}. However, these approaches are generally complex and not easily implementable especially when dealing with many-body quantum systems.

Hence, most studies are restricted to equations that are second order in the system-bath coupling. While being simple, such second-order equations are not devoid of issues. For example, the least approximative Redfield equation can lead to unphysical negative populations~\cite{VRomero1989,ASuarezIOppenheim92,PPechukas94,EGevaERosenman2000,AMCastilloDRReichman15,Strunz20}. The quantum optical master equation (also known as the Lindblad equation) gives an equilibrium reduced density matrix that for a wide class of models is independent of the system-bath coupling strength~\cite{JThingaPHaenggi2012} contrary to the notion of the Hamiltonian of mean force \cite{Jarzynski17, TalknerHaenggi2020}. Moreover, any second-order master equation has issues with its accuracy such that the inaccuracies develop over time leading to a wrong steady state~\cite{TMori2008,FlemingCummings2011}. 

In this letter, we use the asymptotic state of the system to develop a QME that goes beyond these common weak coupling approximations. A similar approach was highly successful to improve on the conventional classical Fokker-Planck equation~\cite{HanggiPRA84}, but its quantum counterpart is still missing. We address this gap and develop a fully quantum formulation that uses the asymptotic state to correctly steer the transient dynamics. In case of a quantum system connected to a single reservoir, we use the generalized canonical distribution also known as the mean-force Gibbs state~\cite{Anders2022}, see Eq.~\eqref{eq:gen_gibbs}, to correct for the transient dynamics and corroborate our findings with the exactly solvable quantum dissipative harmonic oscillator. The generalized canonical distribution incorporates effects of finite system-reservoir coupling giving a solution that also \emph{correctly} captures finite-coupling effects. This \emph{canonically consistent quantum master equation} (CCQME) is as universal and versatile as standard weak-coupling QMEs, making it applicable to a wide range of scenarios ranging from quantum optics, chemical physics, statistical physics, and more recently quantum information and -thermodynamics. 

A key feature of our approach is its simplicity. Even though we obtain solutions that are accurate beyond weak coupling we do not require any additional information than what is needed in a weak-coupling Redfield equation. In other words, we do not require cumbersome fourth-order tensors involving multi-dimensional integrals~\cite{JThingnaWJSheng2014} or elaborate numerical methods~\cite{HEOM89,QUAPI95,RCNazir14,TTM14,FruchtmanLambertGauger2016} that restrict treatable Hilbert space dimensions. We even find accurate agreement with the numerically exact hierarchy equation of motion (HEOM) approach and demonstrate our methods applicability for an interacting many-body open quantum system~\cite{supp}. Moreover, using recent advances in evaluating the asymptotic state of nonequilibrium quantum systems driven by multiple reservoirs~\cite{ThingnaPRE13} or an external drive~\cite{TShirai2016, CaoPRL19}, our approach is easily extendable to study the dynamics of externally or boundary driven quantum systems which we show in the supplemental material \cite{supp} for the damped harmonic oscillator driven by two baths.

\begin{figure}[t]
	\includegraphics{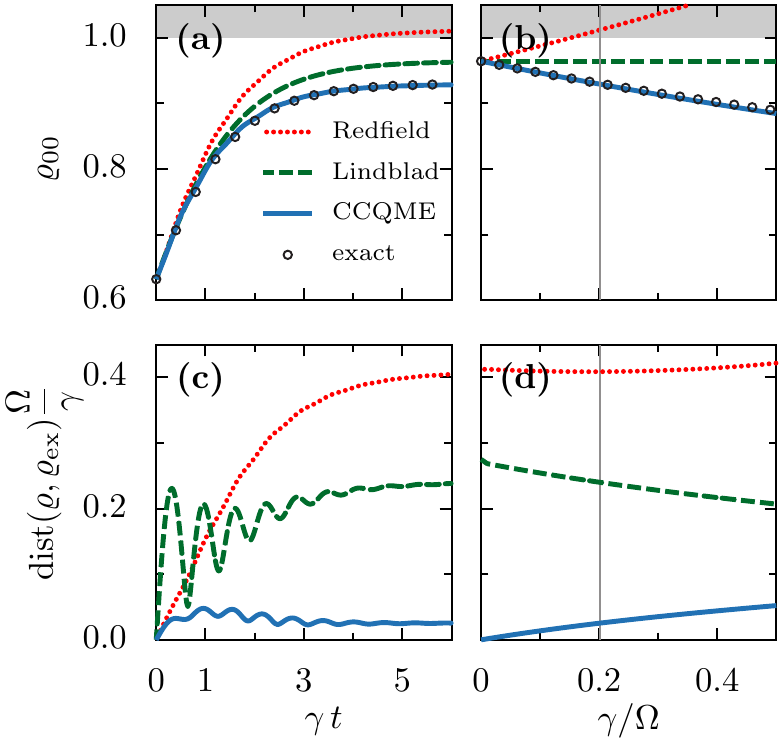}
	\caption{Comparison between the Redfield (dotted red), Lindblad (dashed green), CCQME (solid blue), and the non-perturbative exact solution (open circles) for a harmonic oscillator system with frequency $\Omega$ coupled to a thermal bath with strength $\gamma$. The ground state population dynamics are shown in (a), while (b) depicts the steady-state populations. The gray regions in (a) and (b) indicate unphysical regimes where the ground state populations exceed 1. Deviation of the density matrix to the exact result characterized by the trace distance $\text{dist}(\varrho,\varrho_{\rm ex})$ is shown for the dynamics in (c) and steady state in (d). The system is initiated in $\varrho(0)=\exp[-H_\mathrm{S}/\Omega]/Z_\mathrm{S}$, and we truncate to the $N=60$ lowest levels. The bath temperature is $T/\Omega=0.3$ and has an Ohmic Lorentz-Drude spectral density, $J(\omega) = \gamma \omega/(1+\omega^2/\omega_\mathrm{D}^2)$ with cutoff $\omega_\mathrm{D}/\Omega=5$ and strength $\gamma=0.2$ [(a) and (c)] marked by a gray vertical line in (b) and (d).} 
\label{fig:main}
\end{figure}

\noindent \textit{Preliminaries.--} We consider an autonomous system of interest $H_\mathrm{S}$ coupled to a thermal bath $H_{\mathrm{B}}$ such that the composite Hamiltonian reads $\hat{H}_\mathrm{tot}=H_0 + \hat{H}_\mathrm{int}$ with $H_0 = \hat{H}_\mathrm{S}+\hat{H}_\mathrm{B}$. The system couples to the reservoir via a general interaction $\hat{H}_\mathrm{int}= \hat{S}\otimes\hat{B}$ with $\hat{S}$ ($\hat{B}$) being any hermitian operator acting in the system (bath) Hilbert space. We also address the case of several coupling terms further below in the context of boundary driven systems. Throughout this letter we work in units where $\hbar=k_\mathrm{B}=1$. The composite system evolves uni\-ta\-ri\-ly and we are interested in the dynamics of the reduced density matrix $\varrho = \tr_\mathrm{B}(\varrho_\mathrm{tot}) \equiv  \Lambda_t[ \varrho(0)]$. Under weak-coupling and Born approximations using decoupled initial conditions the dynamical map is given by~\cite{breuerpetruccione} (see also supplemental material~\cite{supp}),
\begin{align} 
 \Lambda_t[\cdot ] &\simeq\Lambda^{0}_{t} \left[ \mathbb{I}[\cdot]+\int_0^t  \mathrm{d}\tau \Lambda^{0}_{-\tau}[ \mathcal{R}_{\tau}[ \Lambda^{0}_{\tau}[\cdot] ] ] \right] . 
\label{eq:lamb-til}
\end{align}
Above, $\Lambda^{0}_{t}[\cdot ] = \exp[-iH_\mathrm{S} t] \cdot  \exp[iH_\mathrm{S} t]$ is the non-interacting evolution superoperator, $\mathbb{I}[X] = X$ the identity superoperator, and $\mathcal{R}_t$ the time-dependent Redfield superoperator~\cite{AGRedfield65} given by
\begin{align} 
	\mathcal{R}_{t}[\cdot]  =  \int_0^{t} \mathrm{d}\tau \left( \left[\tilde{S}(-\tau) \,\cdot, {S}\right]  C(\tau) + \mathrm{H.c.}\right),
\end{align}
where $C(t) = \tr_\mathrm{B} (\exp[iH_\mathrm{B}t]B \exp[-iH_\mathrm{B}t] B \rho_\mathrm{B})$ is the two-point bath correlator with $\rho_\mathrm{B} = \exp[-\beta H_\mathrm{B}]/Z_{\mathrm{B}}$ being the initial density matrix of the bath and $\tilde S(-t) = \Lambda^{0}_{-t}[S]$. To obtain \cref{eq:lamb-til} one uses the Dyson expansion for the unitary time-evolution operator of the composite system to obtain a perturbative series in $H_\mathrm{int}$. Any truncation of this series leads to divergences with respect to time such that at second order the map $\Lambda_t$ diverges linearly in time~\cite{Timm2011,FlemingCummings2011} (see also supplemental material~\cite{supp}). 
This divergence can be avoided if one dynamically coarse grains the map by time integrating the Redfield dissipator in \cref{eq:lamb-til}, which is the starting point of dynamical coarse-graining~\cite{SchallerBrandes2008}. 

Alternatively, in the standard open quantum systems framework one avoids the map and, by taking the time-derivative of \cref{eq:lamb-til}, one obtains a first-order differential equation,
\begin{align}
\partial_t \varrho(t) &=-i \left[H_\mathrm{S}, \varrho(t)\right] +  \mathcal{R}_{t}[ \Lambda^{0}_{t}[\varrho(0)]]. 
\label{eq:redfield_free_evolve}
 \end{align}
The above non-homogeneous first-order differential equation precedes the Redfield equation since the dissipator $\mathcal{R}_t$ acts on the initial state $\varrho(0)$. The differential form ensures that there are no divergences at the second order in the asymptotic limit, these are simply pushed to higher orders~\cite{JThingnaWJSheng2014}. In other words, since the divergence in the map grows linearly with time, differentiating with respect to time eliminates the divergence at the cost of obtaining a differential equation rather than a map. At the level of second order in $H_{\rm int}$, one often replaces
\begin{equation}
\label{eq:approx}
\Lambda_t^{0}[\varrho(0)] \approx \varrho(t).
\end{equation}
As we see from \cref{eq:lamb-til}, the above substitution results in errors at a higher order in system-bath coupling leaving the differential equation correct upto second order, as desired. The resulting QME is known as the time-dependent Redfield equation~\cite{AGRedfield65} that violates complete positivity (CP) and hence is less preferred over the (secular) Lindblad equation~\cite{GLindblad1976, GoriniKossakowski1976,breuerpetruccione,supp}. Despite the lack of CP, a stringent restriction for physical maps~\cite{PPechukas94, sudarshan2005}, the Redfield equation is able to capture finite system-bath coupling effects for which the Lindblad equation is insensitive~\cite{XXuJThinga2019,Strunz20}. Thus, it is only recently that the Redfield equation has gained popularity as a tool to incorporate finite system-bath coupling effects~\cite{JThingaPHaenggi2012,ThingnaPRE13,TShirai2016,Purkayastha22,Alicki22,Trushechkin22}. In \cref{fig:main} we demonstrate this for a dissipative harmonic oscillator, described in detail later in the manuscript. The Redfield equation violates positivity in \cref{fig:main}(a) and (b) and errors build up over time such that the steady state shows finite errors in second order in $H_{\rm int}$ [see \cref{fig:main}(d)].
To resolve these issues, we propose below a scheme that uses the mean--force Gibbs state from equilibrium statistical mechanics~\cite{Jarzynski17,TalknerHaenggi2020,Anders2022} to correct the Redfield equation, specifically improving on the approximation in Eq.~\eqref{eq:approx}. Our approach, despite not being CP, avoids unphysical negative populations for finite coupling strengths (see Fig.~\ref{fig:main}) and ensures that the equilibrium state is mean force Gibbs~\cite{Jarzynski17,TalknerHaenggi2020,Anders2022} upto second order in system-bath coupling. 

\noindent \textit{Canonically consistent quantum master equation (CCQME).--} To start with, we assume $Q_t$ to be the superoperator that generates the second order contributions for the time evolution
\begin{align}
	\varrho(t) &\simeq \Big (\mathbb{I} + Q_t\Big)[\Lambda^{0}_t[\varrho(0)]] \label{eq:eff_dyn_map}.
\end{align}
To find the Markovian master equation generated by this dynamical map we take the formal inverse $(\mathbb{I} + Q_t)^{-1} \simeq (\mathbb{I} -  Q_t)$ and use this relation to replace the freely evolving state on the right hand side of \cref{eq:redfield_free_evolve}. We obtain the QME 
	$\partial_t \varrho(t) =-i \left[H_\mathrm{S}, \varrho(t)\right] +  \mathcal{R}_{t} \left[ (\mathbb{I} - Q_t) [\varrho(t)] \right], $
which involves a fourth-order correction to the Redfield equation.


Instead of following pathological perturbative expansions for the dynamical map that involves divergences in each order separately we make a proposal that can be viewed to combine notions of quantum mechanics and statistical mechanics in a holistic way. While quantum me\-cha\-nics is a dynamical theory for the microscopic degrees of freedom, statistical mechanics describes systems in equilibrium by very few parameters only. Profound mechanisms have been developed to connect both fields, such as eigenstate thermalization hypothesis~\cite{Deutsch_2018}, canonical-~\cite{LebowitzPRL06}, and dynamical~\cite{GemmerPRL09} typicality. For finite coupling, we assume that once the system is coupled to the reservoir the system eventually relaxes to the mean force Gibbs distribution 
\begin{align}
	\lim_{t \rightarrow \infty}\varrho(t) & = \frac{\tr_\mathrm{B}(e^{-\beta H_\mathrm{tot}})}{Z_\mathrm{tot}} \simeq  \Big(\mathbb{I} + \bar{Q} \Big) \left[\varrho_\mathrm{G}\right], 
	\label{eq:gen_gibbs}
\end{align}
with $\varrho_\mathrm{G}=\exp[-\beta H_\mathrm{S}]/{Z_\mathrm{S}}$ being the canonical Gibbs state obtained in the infinitesimally weak-coupling long-time limit of $\varrho(t)$. 
In particular, for finite coupling, the reduced state of the system deviates from the canonical Gibbs distribution and the lowest-order correction $\bar{Q}[\varrho_\mathrm{G}]$ is second order in system-bath coupling.

Based on the similarities between the dynamical map, \cref{eq:eff_dyn_map}, and the equilibrium state, \cref{eq:gen_gibbs}, it becomes evident that $Q_\infty$ can be replaced with $\bar{Q}$. Note here that the superoperators $Q_\infty$ and $\bar{Q}$ are not strictly equivalent. The difference between them stems from the order in which the long-time and weak-coupling limits are performed. In equilibrium statistical mechanics, which leads to the mean-force Gibbs state, we take the infinite-time limit first, followed by the weak-coupling limit to obtain a non-divergent $\bar{Q}$, whereas within the quantum framework we take the weak-coupling limit first and the infinite-time limit later to obtain $Q_{\infty}$ that is divergent. Such non-commutative order of limits leading to different results appears in different areas of physics~\cite{GiulianiVignale2005,Silbey06}.

In problems such as those tackled in this work, 
a sensible (convergent) answer is given when the order of the limits is dictated by statistical mechanics rather than quantum dynamics.  
%
Therefore, we replace $Q_t$ with its 
long-time version given by statistical mechanics, $\bar{Q}$. 
This gives our main result, the CCQME
\begin{align}
	\partial_t \varrho(t) &=-i \left[H_\mathrm{S}, \varrho(t)\right] +  \mathcal{R}_{t} \left[ (\mathbb{I} - \bar{Q}) [\varrho(t)] \right].
	\label{eq:ccqme}
\end{align}
It goes beyond standard second-order treatments and is consistent with statistical mechanics.
Note the following subtlety:
Since \cref{eq:gen_gibbs} only fixes the action of  the superoperator $\bar Q$ on the Gibbs state, the action of  $\bar Q$ on all other states is
in principle undetermined (apart from the fact that $\bar Q$ needs to preserve hermiticity). 
Different CCQMEs\- are, thus, possible that are expected to perform equally well for the steady state, however the transient dynamics depends on the particular choice. 

By construction it is straight-forward to prove [using Eq.~\eqref{eq:gen_gibbs} in Eq.~\eqref{eq:ccqme}] that the steady-state solution of the CCQME is given by the mean force Gibbs state, 
\begin{align}
	0 = 
	-i \Delta \left[\bar Q\left[\varrho_\mathrm{G} \right]\right] 
	+ \mathcal{R}_{\infty} \Bigl[\underbrace{(\mathbb{I} - \bar Q)\Bigl[(\mathbb{I} + \bar Q )}_{\simeq \mathbb{I}} \left[\varrho_\mathrm{G}\right]\Bigr]\Bigr]  ,
	\label{eq:steady_state}
\end{align}
with $\Delta[\cdot] = [\hat H_\mathrm{S}, \cdot]$. The above equation yields the necessary condition $i \Delta [\bar Q[\cdot]] =  \mathcal{R}_{\infty}[\cdot]$, which we later show to be valid. The CCQME does not need to be completely positive but relaxes towards the positive mean-force Gibbs state. Above the terms in the under-brace are approximated to the identity superoperator by ignoring sixth order system-bath coupling strength contributions. For the CCQME, the coupling strength $\gamma$ can be viewed as a parameter that governs the precision of the dynamics. Positivity of the density matrix is ensured only when the strength $\gamma$ is within the limits of the approximations applied (see supplemental material for more details~\cite{supp}).

Using canonical perturbation theory~\cite{JThingaPHaenggi2012}, we show that one choice for the superoperator $\bar Q$ can be expressed as (see supplemental material~\cite{supp}),
\begin{align}
	\begin{aligned}
	\bar Q [\varrho] = &\mathcal P \frac{1}{i\Delta} [\mathcal{R}_{\infty}[\varrho]] \\& +\mathcal P^c \left( \sum_{nl} \left[ \mathcal{L}(L_{nl})[\varrho]  +  |S_{nl}|^2 W''_{ln} \frac{\partial\varrho}{\partial E_n} \right]\right). 
	\end{aligned}
	\label{eq:Q-expl}
\end{align}
Above $\mathcal{P}=\sum_{n\ne m} \mathcal{P}_{nm}$ is the projector into the coherent subspace with $\mathcal{P}_{nm} = \ketbra{n}{n}{\cdot} \ketbra{m}{m}$. 
Moreover, $\ket{n}$ ($E_n$) are the eigenstates (non-degenerate eigenenergies) of $H_\mathrm{S}$ so that the superoperator $\Delta$ is invertible in this subspace, and $\mathcal P^c = \sum_{n} \mathcal{P}_{nn}$ is the projector into the complimentary subspace of $\mathcal{P}$. 
Acting with the superoperator $i \Delta$ on \cref{eq:Q-expl}, the second term vanishes due to the commutator in the superoperator $\Delta$, and  the first term yields $i \Delta[ \bar Q[\cdot]] =  \mathcal{R}_{\infty}[\cdot]$, which satisfies the necessary condition for the steady-state equation above. 
In the complementary subspace for which $\mathcal P^c$ projects to the populations 
we find a Lindblad contribution $\mathcal{L}(L) = L \cdot L^\dagger - \frac{1}{2}\lbrace L^\dagger L , \cdot \rbrace$ with jump operators $L_{nl} =\sqrt{|S_{nl}|^2 V''_{nl}}\ketbra{n}{l}$, 
with  $V''_{nl}$ being the imaginary part of $V(E_{n}-E_l)=V'_{nl}+iV''_{nl}$. Here $V(E)= \partial W(E)/\partial E$ is defined using the Fourier-Laplace transform of the bath correlator $W(E)= \int_0^\infty \mathrm dt\,  C(t) \exp[-i E t]$ used in the Redfield equation. The function $W(E_n-E_l)=W'_{nl}+iW''_{nl}$ contains real and imaginary parts typically referred to as rates and lamb shifts, respectively. 
On the other hand the last term in Eq.~\eqref{eq:Q-expl} can be obtained as~\cite{supp}, 
\begin{align}
	\frac{\partial \varrho}{\partial E_n} = \frac{\sum_{l \ne n} |S_{nl}|^2 \, [ V'_{nl}\, p_l +  V'_{ln}\, p_n] }{\sum_{l\ne n} |S_{ln}|^2\, W'_{ln} } \ketbra{n}{n},
	\label{eq:formal_deriv}
\end{align} 
where $p_n = \matrixel{n}{\varrho}{n}$ are the populations. For the canonical Gibbs state this term simplifies to $\partial \varrho_\mathrm{G}/\partial E_n = -\beta \mathcal P_{nn}[\varrho_\mathrm{G}]$.

\noindent \textit{Benchmark with exact dynamics.--} We corroborate our method with the exactly solvable dynamics for the damped harmonic oscillator whose total system-bath Hamiltonian \cite{CaldeiraLegget83} reads
		$\hat{H}_\mathrm{tot} =H_\mathrm{S} + \sum_k^\infty [ {\hat{p}_k^2}/({2m_k}) + {m_k \omega_{k}^2} (\hat{q}_k - {c_k}{m_k^{-1} \omega_k^{-2}} \hat{q} )^2/2 ].$
In units where the particle mass is set to one we have $H_\mathrm{S} =  \hat{p}^2/2 + \Omega^2\, \hat{q}^2/2 $. For a bath in thermal equilibrium that is factorized initially from the state of the oscillator one obtains the exact (non-perturbative in $\gamma$) QME~\cite{HaakeReibold85,HuPazZhang1992,KarrleinGrabert97,Paz94,JPiilo2004},
\begin{align*}
	\begin{split}
	\partial_t \varrho_\mathrm{ex}(t)=& -\frac{i}{2} [\hat{p}^2 + \gamma_q(t)\hat{q}^2, \varrho_\mathrm{ex}(t)] - D_p(t) [q,[q,\hat{\varrho}_\mathrm{ex}(t)]] \\ 
	&- \frac{i}{2} \gamma_p(t) [\hat{q},\{\hat{p},\hat{\varrho}_\mathrm{ex}(t)\}]+D_q(t) [\hat{q},[\hat{p},\hat{\varrho}_\mathrm{ex}(t)]].
	\end{split}
\end{align*}
Here the damping coefficient $\gamma_p(t)$ is derived from the system correlation and the diffusive coefficients $D_q(t)$ and $D_p(t)$ depend on the bath correlation. A detailed discussion of the parameters can be found in Refs.~\cite{supp,HaakeReibold85,HuPazZhang1992,KarrleinGrabert97}. 

In \cref{fig:main} we benchmark the dynamics [(a) and (c)] and the steady state [(b) and (d)] obtained via the Redfield equation, the Lindblad equation, and the CCQME with the exact solution. For all simulations we assume that the bath correlations decay faster than the time scale of the dynamics leading to an autonomous generator (see~\cite{supp} for the spin-boson model solved with the full time-dependent generator). For strong coupling $\gamma/\Omega\simeq0.2$ and low bath temperature $T/\Omega=0.3$ we observe a breakdown of the Redfield theory as the ground state population exceeds one in \cref{fig:main}(a). We quantify the deviation from the exact result, i.e., error, with the trace distance $\mathrm{dist}(\varrho,\varrho_\mathrm{ex})=(1/2) \tr (\sqrt{ (\varrho(t) - \varrho_\mathrm{ex}(t))^2 })$, a metric bounded by one. 
At small temperatures the trace distance is nearly equivalent to the percentage error in the ground-state population, i.e., the maximum trace distance of $0.08$ (for $\gamma = 0.2\Omega$) for the Redfield case in Fig.~\ref{fig:main}(c) represents a maximum error of $\approx 8\%$. The Lindblad equation shows large oscillations in the trace distance ($\approx 5\%$) but does not suffer from unphysical solutions. The CCQME reproduces the ground state population for all times in \cref{fig:main}(a) and shows small deviations ($< 1\% ~\forall t$) from the exact solution in \cref{fig:main}(c) improving upon the Redfield and Lindblad solutions in the strong-coupling and low-temperature regime.

\begin{figure}[t]
	\includegraphics{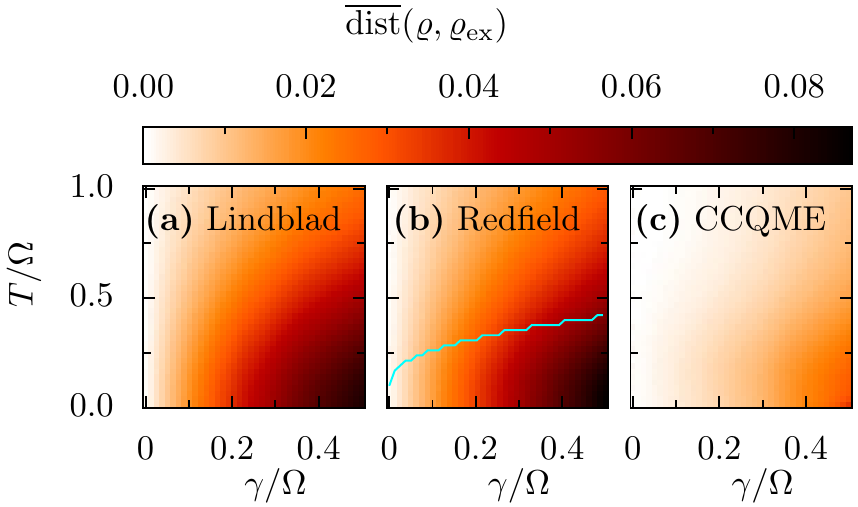}
	\caption{Time averaged trace distance ($\overline{\mathrm{dist}}(\varrho,\varrho_\mathrm{ex}) = \tau_\mathrm{R}^{-1}\int_0^{\tau_\mathrm{R}}\mathrm{d}t\,\mathrm{dist}(\varrho,\varrho_\mathrm{ex})$) over relaxation time $\tau_\mathrm{R} = 2/\gamma$ 
	as a function of temperature and coupling strength for the Lindblad equation (a), Redfield equation (b), and CCQME (c). The Redfield steady-state solution violates positivity in the parameter regime below the solid teal line. Other parameters are same as Fig.~\ref{fig:main}}
	\label{fig:grid}
\end{figure}

In equilibrium, finite coupling leads to an effective higher temperature such that the ground state population decreases as a function of the coupling strength. In \cref{fig:main}(b) the CCQME accurately reproduces this trend and also perfectly matches with the exact curve in the strong coupling regime. On the one hand, the Lindblad solution remains independent of the system-bath coupling $\gamma$ yielding the infinitesimally weak coupling result of $\varrho_\mathrm{G} = \exp[-\beta H_{\rm S}]/Z_{\rm S}$. 
On the other hand, the Redfield not only gives an incorrect ground state population but also erroneously predicts that as the system-bath coupling strength increases the system cools down
. At moderate values of the coupling, the ground state population $\varrho_{00}>1$ implying that the excited states have unphysical negative populations. Since both the exact steady-state solution and the approximate ones from QMEs contain all powers of $\gamma$, we would like to quantify how much do the QMEs deviate from the exact solution at the lowest non-trivial order, i.e., second order. This deviation in the solution can be quantified by dividing the trace distance by the coupling strength and numerically approaching the limit of $\gamma/\Omega \to 0$ [see \cref{fig:main}(d)]. If the QME and the exact solutions match at second-order $O(\gamma)$, this measure approaches zero as $\gamma/\Omega \to 0$ as seen for the CCQME [solid blue line in \cref{fig:main}(b)]. The other QMEs give a finite deviation indicating discrepancies at $O(\gamma)$.


To get a complete overview of the deviations even in the dynamics for the entire parameter space, in \cref{fig:grid} we calculate the time-averaged trace distance to the exact result for the entire relaxation process as a function temperature and coupling strength initiating the system in an out-of-equilibrium initial state. For strong coupling the Lindblad and Redfield equations are only valid for high bath temperatures, i.e., deep classical limit. Contrastingly, the CCQME leads to a more accurate result in the full parameter regime. Importantly, the CCQME solution gives positive density matrices in the entire parameter range in sharp contrast to the Redfield that fails at low temperatures [below solid teal line in~\cref{fig:grid}(b)]. 

\noindent \emph{Quantum Transport.--}
The CCQME can also be implemented for transport setups~\cite{Transport2,Transport3} wherein the system is driven by several independent baths $H_\mathrm{B}=\sum_i H_\mathrm{B}^i$ that couple via the interaction Hamiltonian $H_\mathrm{int}=\sum_i S^i \otimes B^i$. At second order of the interaction, the Redfield and Lindblad equation is obtained by adding the dissipative superoperators for the individual baths. However beyond second order cross correlations build up \cite{CaoJCP, JThingnaWJSheng2014}, which occur naturally in the boundary driven CCQME 
\begin{align}
	\partial_t \varrho(t) = -i [H_\mathrm{S},\varrho(t)] + \sum_i \mathcal{R}_t^i [(\mathbb{I} - \sum_j \bar Q^j)[\varrho(t)]],
	\label{eq:boundaryCCQME}
\end{align}
by the products of superoperators for different baths. We elucidate this idea further by studying a boundary driven harmonic oscillator and corroborating the CCQME with exact results in the supplemental material~\cite{supp}. 

\noindent \emph{Summary.--}
In this work, we proposed a QME that corrects the standard Born-Markov equations (Redfield or Lindblad) incorporating effects of higher order system-bath coupling. The CCQME draws inspiration from the statistical mean-force Gibbs state and correctly steers the dynamics of a quantum system coupled with finite strength to a reservoir. By construction it yields the exact equilibrium state up to second order of the coupling and significantly improves the dynamics as compared to the Redfield or Lindblad equation, which we explicitly demonstrate for the damped harmonic oscillator. Despite not being completely positive our approach does not suffer from negative solutions. The CCQME is not only accurate but also easy to implement since it requires no additional information as compared to the Redfield equation. 
Moreover it is model independent and could shed light onto finding lowest order effects due to the presence of system-bath coupling in the emerging field of strong-coupling quantum thermodynamics~\cite{EspositoPRB15,Jarzynski17,PhilPRE17,NewmanPRE17,TalknerHaenggi2020} or in dissipative quantum many-body systems as demonstrated in the supplemental material using an Ising chain~\cite{supp}. The CCQME could also be extended to study transport through systems strongly connected to multiple reservoirs aiding the field of strong-coupling quantum transport~\cite{RN2,Dvira21,RN3,PolettiSchaller2021}.



\begin{acknowledgments}
We thank Peter H\"anggi, Peter Talkner, Walter Strunz, André Eckardt, and Gernot Schaller for helpful discussions.
This research was supported by the Institute for Basic Science in South Korea (IBS-R024-Y2) and by the German Research Foundation (DFG) via the Collaborative Research Center SFB910 under project number 163436311.
\end{acknowledgments}

\bibliography{ref}

\clearpage

\title{Supplemental Material for \\ "Canonically consistent quantum master equation"}

\maketitle

\setcounter{equation}{0}
\makeatletter
\renewcommand{\theequation}{A\arabic{equation}}

\subsection{Dyson series expansion of the dynamical map - Divergences at each order}
We consider a general open quantum system whose total Hamiltonian
\begin{align}
	H_\mathrm{tot} = \underbrace{H_\mathrm{S} + H_\mathrm{B}}_{H_0} +  \underbrace{ S \otimes B}_{H_\mathrm{int}},
\end{align}
with $H_\mathrm{S}$ and $S$ being operators acting on the system Hilbert space, and $H_\mathrm{B}$ and $B$ acting on the Hilbert space of the bath.
Additionally, we assume a weak interaction $H_\mathrm{int} $ (we will later refine this requirement).

Using Kubo's identity,
\begin{align}
	\mathrm{e}^{\mu (A+B)} = \mathrm{e}^{\mu A} \left( \mathbb{I} + \int_0^\mu \mathrm{d}\nu \mathrm{e}^{-\nu A} B   \mathrm{e}^{\nu (A+B)} \right),
	\label{eq:Kubo}
\end{align}
we expand the total time evolution operator $U_{\rm tot}(t,0) = \exp[-it(H_0+H_\mathrm{int})]$ upto second order in $H_{\rm int}$ yielding,
\begin{align}
	\begin{split}
		&\mathrm{e}^{-it H_\mathrm{tot}} \simeq \mathrm{e}^{-it H_0} \left[\mathbb{I} -i  \int \tilde{H}_\mathrm{int}(t_1)  \right.\\
		&  \left. -  \iint^{t_1}   \tilde{H}_\mathrm{int}(t_1) \tilde{H}_\mathrm{int}(t_2) \right].
	\end{split}
	\label{eq:timev-exact-dyson}
\end{align}
Above the symbol $\int \coloneqq \int_0^t \mathrm{d}t_1$ and $\iint^x \coloneqq \int_0^t \mathrm{d}t_1  \int_0^{x} \mathrm{d}t_2$. The operator $\tilde{H}_{\rm int}(\tau) = \exp[i H_{0}\tau] H_{\rm int} \exp[-i H_0 \tau] $ is defined in the interaction picture and throughout we will set $\hbar=k_\mathrm{B} = 1$. Thus, the total density matrix
\begin{align}
	\varrho_\mathrm{tot}(t)= \mathrm{e}^{-it(H_0+H_\mathrm{int})} \varrho_\mathrm{tot}(0) \mathrm{e}^{it(H_0+H_\mathrm{int})}.
	\label{eq:full-dyn-exact}
\end{align}
transforms as,
\begin{align}
	\begin{split}
		\mathrm{e}^{it H_0} &\varrho_\mathrm{tot}(t)  \mathrm{e}^{-it H_0}\simeq  \varrho_\mathrm{tot}(0)  -i  \int [ \tilde{H}_\mathrm{int}(t_1), \varrho_\mathrm{tot}(0)] \\
		&- \iint^{t_1}  \left( \tilde{H}_\mathrm{int}(t_1) \tilde{H}_\mathrm{int}(t_2) \varrho_\mathrm{tot}(0)  + \mathrm{H.c.}\right)\\
		&+ \iint^t \tilde{H}_\mathrm{int}(t_1) \varrho_\mathrm{tot}(0) \tilde{H}_\mathrm{int}(t_2).
	\end{split}
	\label{eq:timev-dyson-secondord}
\end{align}
The above is the Dyson expansion up to second order in $H_\mathrm{int}$.
An alternative derivation of this expression can be obtained from a Magnus expansion up to second order in the interaction picture, followed by an expansion of the involved exponential up to second order. 

In the following it is convenient to work in the interaction picture. To obtain the dynamical map $\tilde \varrho(t) = \tilde \Lambda_t[\tilde \varrho(0)]$ for the reduced dynamics $\tilde \varrho(t) = \mathrm{Tr}_\mathrm{B}(\tilde \varrho_\mathrm{tot}(t))$, we assume factorized initial conditions $\tilde \varrho_\mathrm{tot}(0)=\tilde \varrho(0) \otimes \tilde \varrho_\mathrm{B}(0)$ and trace over the bath degrees of freedom to find
\begin{align}
	\begin{split}
		\tilde \varrho(t) &\simeq  \tilde \varrho(0) -i   \int \left[ \tilde{S}(t_1) \langle\tilde{B}(t_1) \rangle, \tilde \varrho(0)  \right] \\
		&+  \iint^{t_1} \left( \left[\tilde{S}(t_2) \tilde \varrho(0), \tilde{S}(t_1) \right]  \langle\tilde{B}(t_1)\tilde{B}(t_2) \rangle + \mathrm{H.c.}\right).
	\end{split}
	\label{eq:def-dyn-til}
\end{align}
Herein $\langle \cdot \rangle = \mathrm{Tr}_\mathrm{B}( \cdot \varrho_\mathrm{B}(0))$ denotes the bath average in the unperturbed (or ``free'') evolution of the bath. Further, we assume that the bath is stationary, therefore the bath correlation function $C(t_1-t_2) =  \langle\tilde{B}(t_1)\tilde{B}(t_2) \rangle $ will only depend on the time difference and not the absolute times $t_1, t_2$. Also, as in the usual convention of Born-Markov, the bath operators are chosen such that $\langle\tilde{B}(t_1) \rangle = 0$ (which can always be achieved by a redefinition of the $B$ operators and $H_\mathrm{S}$). Now by transforming $t_2 \rightarrow t_1-t_2$ in Eq.~\eqref{eq:def-dyn-til}, one obtains
\begin{align}	
	\begin{aligned}
		\tilde \varrho(t)
		&\simeq \tilde \varrho(0) + \int e^{it_1 H_\mathrm{S}} \mathcal{R}_{t_1} [e^{-it_1 H_\mathrm{S}} \tilde \varrho(0) e^{it_1H_\mathrm{S}}] e^{-it_1 H_\mathrm{S}},
		\label{eq:app_dyson_redfield}
	\end{aligned}
\end{align}
with $\mathcal{R}_{t_1}[\cdot] = \int_0^{t_1} \mathrm dt_2 [\tilde S(-t_2) \cdot,S] C(t_2) + \mathrm{H.c.}$ being the Redfield superoperator, which is the starting point for the derivation of a time local quantum master equation discussed in the main text. 

From \cref{eq:app_dyson_redfield} one identifies the dynamical map as $\tilde \Lambda_t [\cdot]\simeq \mathbb{I}+\int_0^t  \mathrm{d}t_1\, \Lambda^{0}_{-t_1}[\mathcal{R}_{t_1}[ \Lambda^{0}_{t_1}[\cdot] ]]$ with $\Lambda_t^{0}[\cdot] = e^{-itH_\mathrm{S}} \cdot e^{itH_\mathrm{S}}$ being the free evolution. However problems arise for the populations since in the asymptotic limit when the Redfield superoperator $\mathcal{R}_{t_1}$ is essentially time independent, the integral diverges linearly with time. This is seen explicitly for an initial state that is diagonal in $H_\mathrm{S}$ for which the dynamical map reads
\begin{align}
	\begin{aligned}
		\mathcal P^c \tilde \Lambda_t \tilde \varrho(0) &\simeq\tilde \varrho(0) \\ &+ \mathcal P^c \Big( \int  (t-\tau) C(\tau)   \left[ \tilde S(-\tau) \tilde \varrho(0), S \right]  \Big) + \mathrm{H.c.} ,
	\end{aligned}
\end{align}
which is obtained from \cref{eq:def-dyn-til} by choosing the new integration variables $\tau = t_1 - t_2$ and $\tau' = t_1 + t_2$ and $\mathcal P^c = \sum_n \ketbra{n}{n}{\cdot}\ketbra{n}{n}$ is the projector to the populations, where $\ket{n}$ are the eigenstates of $H_\mathrm{S}$. Clearly the dynamical map obtained from the Dyson series expansion is inadequate in describing the long-time dynamics as it linearly diverges with $t$. In fact the series expansion involves divergences in each order separately.


\subsection{Secular Lindblad (or quantum optical) master equation }
Combining Eqs.~(3) and~(4) of the main text, we obtain the Redfield quantum master equation 
\begin{align}
	\partial_t \varrho(t) &=-i \left[H_\mathrm{S}, \varrho(t)\right] +  \mathcal{R}_{t}[ \varrho(t)]. 
\end{align}
From this, a Lindblad form can be found in the ultraweak coupling limit where the system--bath coupling strength is weak when compared to all energy splittings $E_{n} - E_m$, $n\neq m$, in the system. By performing the standard secular (or rotating-wave) approximation, we find \cite{breuerpetruccione}
\begin{align}
	\begin{split}
		\partial_t \varrho(t) &=-i \left[H_\mathrm{S} + H^{\mathrm{LS}}, \varrho(t)\right] \\ & + \sum_{nm} \vert S_{nm} \vert^2\,  2W'_{nm}  \mathcal{L}(\ketbra{n}{m})[ \varrho(t)]. 
	\end{split}
\end{align}
with Lamb shift 
\begin{align}
	H^\mathrm{LS} = \sum\limits_{nm} |S_{nm}|^2 W''_{nm} \ketbra{m}{m}.
\end{align}
This is the secular Lindblad (or quantum optical) master equation.

\subsection{Canonical perturbation theory for the mean force Gibbs state} 
Following the eigenstate thermalization hypothesis, in thermal equilibrium we expect the total system-bath composite to thermalize with inverse temperature $\beta$ (determined by the mean energy). By tracing out the bath degrees of freedom, the reduced state of the system is given by the mean force Gibbs state
\begin{align}
	\lim_{t \rightarrow \infty}\varrho(t) & = \frac{\mathrm{Tr}_\mathrm{B}(e^{-\beta H_\mathrm{tot}})}{Z_\mathrm{tot}},
	\label{eq:app_gen_gibbs}
\end{align}
with $\hat H_\mathrm{tot} = \hat H_0 +  \hat H_\mathrm{int}$ where $\hat H_0 = \hat H_\mathrm{S} + \hat H_\mathrm{B}$ is the Hamiltonian of the system and bath respectively and $\hat H_\mathrm{int}$ is the system--bath interaction. Note that  \cref{eq:app_gen_gibbs}  allows for the identification of the Hamiltonian of mean force as in Refs.~\cite{TalknerHaenggi2020,Anders2022}. 
Here, however, we want to extract the correction to the Gibbs state perturbatively in the interaction Hamiltonian.
By again making use of the Kubo identity, \cref{eq:Kubo}, 
we find to second order,
\begin{align}
	\begin{split}
		e^{-\beta \hat H_\mathrm{tot}} &\simeq e^{-\beta \hat H_0} \Bigl[ \hat \mathbb{I} +  \int_0^\beta \mathrm d\mu\  \tilde {\hat H}_\mathrm{int}(-i\mu) \\
		&+ \int_0^\beta \mathrm d\mu\int_0^\mu \mathrm d\xi \ \tilde {\hat H}_\mathrm{int}(-i\mu)  \tilde {\hat H}_\mathrm{int}(-i\xi)\Bigr] 
	\end{split}
\end{align}
with freely evolving interaction-picture operators in imaginary time $\tilde {\hat H}_\mathrm{int}(-i\mu) = e^{\mu \hat H_0} \hat H_\mathrm{int} e^{-\mu \hat H_0}$. By assuming a bath at equilibrium and by taking the partial trace only even orders of the expansion contribute to the mean force Gibbs state. For a system bath coupling Hamiltonian of the form $\hat{H}_\mathrm{int} = \hat S \otimes \hat B$ this is given by,
\begin{align}
	\begin{split}
		&\mathrm{Tr}_\mathrm{B} (e^{-\beta \hat H_\mathrm{tot}}) =  e^{-\beta \hat H_\mathrm{S}} \ Z_\mathrm{B} \ \Bigl[ \mathbb{I} + \\ 
		& \int_0^\beta \mathrm d\mu\int_0^\mu \mathrm d\xi \ \tilde {\hat S}(-i\mu)  \tilde {\hat S}(-i\xi) C(-i(\mu-\xi)) \Bigr],
	\end{split}
\end{align}
with bath correlation ${C(-i(\mu-\xi))} = {\mathrm{Tr}_\mathrm{B}(\varrho_\mathrm{B} \tilde{\hat B}(-i\mu) \tilde{\hat B}(-i\xi))}$ evaluated at imaginary time. Apart from the system-bath coupling, the Caldeira Leggett model in \cref{eq:appHtot} also considers a renormalization Hamiltonian that renormalizes the frequency shift caused by the reservoir. For an Ohmic bath with Drude cutoff it takes the form $H_\mathrm{RN} = \gamma\omega_\mathrm{D} S^2/2$~\cite{JThingaPHaenggi2012} and modifies the bath correlation, $C(-i \lambda) \rightarrow C(-i \lambda) - \gamma\omega_\mathrm{D} \delta(\lambda)/2$. We arrive at the second order expansion of the mean force Gibbs state,
\begin{align}
	\begin{aligned}
		\frac{\mathrm{Tr}_\mathrm{B}(e^{-\beta H_\mathrm{tot}})}{Z_\mathrm{tot}} \simeq &\Big(\hat \mathbb{I} + \bar{Q} \Big) \left[\frac{e^{-\beta H_\mathrm{S}}}{Z_\mathrm{S}}\right] \\ &- \frac{e^{-\beta H_\mathrm{S}}}{Z_\mathrm{S}} \mathrm{Tr}_\mathrm{S}\,  \bar{Q}\left[\frac{e^{-\beta H_\mathrm{S}}}{Z_\mathrm{S}}\right] 
	\end{aligned}
	\label{eq:app_gen_gibbs_nonlinear}
\end{align}
where we define the action of the superoperator 
\begin{align}
	\begin{aligned}
		\bar{Q} &\left[\frac{e^{-\beta H_\mathrm{S}}}{Z_\mathrm{S}}\right]  = \\& \frac{e^{-\beta H_\mathrm{S}}}{Z_\mathrm{S}}  \int_0^\beta \mathrm d\mu\int_0^\mu \mathrm d\xi \ \tilde {\hat S}(-i\mu)  \tilde {\hat S}(-i\xi) C(-i(\mu-\xi)) 
	\end{aligned}
\end{align}
on the canonical Gibbs state.
The last term in \cref{eq:app_gen_gibbs_nonlinear} that is quadratic in the canonical Gibbs state is a consequence of the normalization. 
In the dynamical theory for which we propose a canonically consistent master equation, which by construction guarantees trace preservation, this non-linear term may be dropped.

In the following we explicitly calculate the action of the operator above and express it in terms of the Redfield dissipator. By rotating the coordinate system and introducing the independent integration variables $s\equiv \mu + \xi$ and $u\equiv \mu - \xi$ we obtain
\begin{align}
	\begin{aligned}
		\bar{Q} \left[\frac{e^{-\beta H_\mathrm{S}}}{Z_\mathrm{S}}\right] =&\frac{e^{-\beta H_\mathrm{S}}}{Z_\mathrm{S}} \int_0^\beta \mathrm du\ C(-iu) \\ & \int_{u/2}^{\beta-u/2} \mathrm ds\ e^{s\hat H_\mathrm{S}} \tilde{\hat S}(-iu/2)\tilde{\hat S}(iu/2) e^{-s\hat H_\mathrm{S}},
	\end{aligned}
\end{align}
in which we can solve the $s$-integration independently. The integral is qualitatively different for either coherences and populations. To keep the notation simple we introduce the projection operators
\begin{align}
	\mathcal{P}=\sum_{n\ne m} \ketbra{n}{n}{\cdot}\ketbra{m}{m}, 
\end{align}
which projects to the coherent subspace where $\Delta=[\hat H_\mathrm{S},\cdot]$ is invertible, and the complementary space
\begin{align}
	\mathcal P^c = \sum_{n} \ketbra{n}{n}{\cdot}\ketbra{n}{n}, 
\end{align}
for the populations. In both subspaces the $s$-integral can be evaluated as,
\begin{align}
	\mathcal P \int \mathrm{d}s \ e^{s\hat H_\mathrm{S}} \cdot e^{-s\hat H_\mathrm{S}} &= \mathcal P\ \frac{1}{\Delta} \ e^{s\hat H_\mathrm{S}} \cdot e^{-s\hat H_\mathrm{S}} , \\
	\mathcal P^c \int \mathrm{d}s\ e^{sH} \cdot e^{-sH} &= \mathcal P^c\ s .
\end{align}
Thus, we obtain the simplified action of the $\bar{Q}$ superoperator as, 
\begin{widetext}
	\begin{align}
		\begin{aligned}
			\bar{Q} \left[\frac{e^{-\beta H_\mathrm{S}}}{Z_\mathrm{S}}\right] = &\mathcal P\ \frac{1}{\Delta} \left[ S \left(\int_0^\beta \mathrm du\, C(-iu) e^{-uH_\mathrm{S}} S e^{uH_\mathrm{S}} \right) \frac{e^{-\beta H_\mathrm{S}}}{Z_\mathrm{S}} - \frac{e^{-\beta H_\mathrm{S}}}{Z_\mathrm{S}} \left(\int_0^\beta \mathrm du\, C(-iu) e^{-uH_\mathrm{S}} S e^{uH_\mathrm{S}} \right)^\dagger  S\right]  \\
			+&\mathcal P^c\   \frac{e^{-\beta H_\mathrm{S}}}{Z_\mathrm{S}} S \int_{0}^{\beta} \mathrm{d}u\, C(-iu) (\beta-u) e^{-uH_\mathrm{S}}Se^{uH_\mathrm{S}}.
		\end{aligned}
		\label{eq:CPTQ}
	\end{align}
\end{widetext}
Integrals of the form above are well known from the Redfield theory. In fact it turns out that the first line can be rewritten with the Redfield superoperator. To this end we make use of the identity \cite{JThingaPHaenggi2012},
\begin{align}
	- \int_0^\beta \mathrm du\, C(-iu) e^{-uE} = W''(E) + e^{- \beta E} W''(-E)
	\label{eq:app_imag_time_identity}
\end{align}
with $W''(x)$ being the imaginary part of the Fourier-Laplace transform of the bath correlation function,
\begin{align}
	W(E) = \int_0^\infty \mathrm dt\, C(t) e^{-i E t}.
\end{align}
The real part of the above integral is typically referred to as the rates whereas the imaginary part is known as the Lamb shift. With this we can rewrite the second order contribution as
\begin{align}
	\begin{aligned}
		&\bar{Q} \left[\frac{e^{-\beta H_\mathrm{S}}}{Z_\mathrm{S}}\right] = \mathcal P\ \frac{1}{i\Delta} \mathcal{R}_{\infty} \left[ \frac{e^{-\beta H_\mathrm{S}}}{Z_\mathrm{S}} \right] \\ &+ \mathcal P^c\   \frac{e^{-\beta H_\mathrm{S}}}{Z_\mathrm{S}} S \int_{0}^{\beta} \mathrm{d}u\, C(-iu) (\beta-u) e^{-uH_\mathrm{S}}Se^{uH_\mathrm{S}},
	\end{aligned}
	\label{eq:fullRed}
\end{align}
with the static Redfield superoperator 
\begin{align}
	\mathcal{R}_{\infty}\left[\varrho\right] = -S \mathbb{S} \varrho + S \varrho \mathbb{S}^\dagger - \varrho \mathbb{S}^\dagger S + \mathbb{S} \varrho S.
\end{align}
and the convolution operator
\begin{align}
	\mathbb{S} = \int_0^\infty \mathrm dt\, C(t) e^{-iH_\mathrm{S} t} S e^{iH_\mathrm{S} t}. 
\end{align}
When $\mathbb{S}$ is expressed in the eigenbasis of $H_\mathrm{S}$ it simplifies as $\matrixel{n}{\mathbb{S}}{m}= S_{nm} W(\Delta_{nm})$ with $S_{nm}= \matrixel{n}{S}{m}$ and eigenenergy difference $\Delta_{nm} = E_n - E_m$. Note, the identity \cref{eq:app_imag_time_identity} contains only the imaginary parts of the Fourier-Laplace transform of the bath correlator. However, in Eq.~\eqref{eq:fullRed} we took the full Redfield superoperator including the real part, i.e., the Redfield rates. Since the Redfield superoperator in Eq.~\eqref{eq:fullRed} acts on the canonical Gibbs state, the additional parts due to the rates sum upto zero as they follow detailed balance. In other words, Eqs.~\eqref{eq:CPTQ} and~\eqref{eq:fullRed} are equal to each other as a consequence of the canonical Gibbs state obeying detailed balance.

We also want to infer the superoperator in the orthogonal subspace for the populations. 
These terms arise in the expansion above as integrals of the form
\begin{align}
	- \int_0^\beta \mathrm du\, C(-iu)\, u e^{-ux} = \frac{\partial}{\partial x} \left( W''(x) + e^{-x \beta} W''(-x)\right).
\end{align}
By making use of this identity in the eigenbasis of $H_\mathrm{S}$ the populations for the second order contribution are given by \cite{JThingaPHaenggi2012}
\begin{widetext}
	\begin{align}
		\matrixel{n}{\bar{Q} \left[\frac{e^{-\beta H_\mathrm{S}}}{Z_\mathrm{S}}\right] }{n} = \sum_l |S_{nl}|^2 \left[ V''(\Delta_{nl}) \frac{e^{-\beta E_l}}{Z_\mathrm{S}} - V''(\Delta_{ln}) \frac{e^{-\beta E_n}}{Z_\mathrm{S}}  + W''(\Delta_{ln}) \frac{\partial}{\partial  E_n} \frac{e^{-\beta E_n}}{Z_\mathrm{S}}  \right].
	\end{align}
\end{widetext}
Here the first two terms describe a Pauli rate equation that can be represented by a Lindblad equation with rates given by the derivatives of the Redfield Lamb shifts $V''(x) = \partial W''(x)/\partial x$. In other words, the Lamb shifts that are typically ignored play an important role as they appear as rates at a higher order. The last term describes yet another contribution that acts as a derivative onto the state itself. Instead of writing the derivative only for the equilibrium state as $\partial/\partial  E_n e^{-\beta  E_n}/Z_\mathrm{S} = -\beta e^{-\beta  E_n}/Z_\mathrm{S}$ we find a much better performance of the CCQME if we infer the derivative operator as in reference \cite{JThingaPHaenggi2012}. To do so we formally solve the Pauli rate equation for the zeroth order steady state populations $p_{n} = e^{-\beta  E_n}/Z_\mathrm{S}$ 
\begin{align}
	p_n = \frac{\sum_{l \ne n} |S_{nl}|^2\, W'(\Delta_{nl})\ p_l }{\sum_{l\ne n} |S_{ln}|^2\, W'(\Delta_{ln}) },
\end{align}
and take the derivative on both sides to find 
\begin{align}
	\frac{\partial p_n}{\partial  E_n} = \frac{\sum_{l \ne n} |S_{nl}|^2 \, [ V'(\Delta_{nl})\, p_l +  V'(\Delta_{ln})\, p_n] }{\sum_{l\ne n} |S_{ln}|^2\, W'(\Delta_{ln}) }
	\label{eq:app_formal_deriv}
\end{align}
with $V'(x) = \partial W'(x)/\partial x$. To conclude this section we identify the superoperator $\bar Q$ for the CCQME from the mean force Gibbs state that is easily obtained from the Redfield rates/Lamb shifts and their derivatives according to
\begin{widetext}
	\begin{align}
		\bar Q [\varrho] = \mathcal P \frac{1}{i\Delta} \mathcal{R}_{\infty}[\varrho] + \mathcal P^c \left( \sum_{nl} |S_{nl}|^2 \left[ V''(\Delta_{nl}) \mathcal{L}(\ketbra{n}{l})[\varrho]  + W''(\Delta_{ln})\ket{n} \frac{\partial }{\partial  E_n}\matrixel{n}{\varrho}{n} \bra{n}\right]\right)
		\label{eq:app_Q2_redfield}
	\end{align}
\end{widetext}
with Lindblad operator $\mathcal{L}(L) = L \cdot L^\dagger - \frac{1}{2}\lbrace L^\dagger L , \cdot \rbrace$ and the formal derivative \cref{eq:app_formal_deriv} for the populations.

\subsection{Exact master equation for damped harmonic oscillator}
In order to corroborate our approach, we test our method against the exact solution of the damped harmonic oscillator, whose position is coupled to a continuum of oscillator modes. The total system-bath Hamiltonian is given by the Caldeira Leggett model \cite{CaldeiraLegget83},
\begin{align}
	\begin{aligned}
		\hat{H}_\mathrm{tot} &= \frac{\hat{p}^2}{2M} + \frac{M \Omega^2}{2} \hat{q}^2 \\&+ \sum_k^\infty \Bigg[ \frac{\hat{p}_k^2}{2m_k} + \frac{m_k \omega_{k}^2}{2} \Big(\hat{q}_k - \frac{c_k}{m_k \omega_k^2} \hat{q} \Big)^2 \Bigg],
	\end{aligned}
	\label{eq:appHtot}
\end{align}
with position $\hat{q}$ ($q_k$) and momentum $\hat{p}$ ($p_k$) of the central (bath) oscillator(s). In the following we work in units where we set the central oscillator mass to one, i.e.\ $M=1$. This model has been extensively studied in references \cite{HaakeReibold85,HuPazZhang1992,KarrleinGrabert97}. Here we are interested in an effective description of the central oscillator that is exact and does not rely on perturbative expansions in system-bath coupling. For this we assume an initial product state of system and bath and that the bath initially is in thermal equilibrium with inverse temperature $\beta$. Under this conditions it has been shown~\cite{KarrleinGrabert97} that the dynamics of the central oscillator follows the exact master equation
\begin{widetext}
	\begin{align}
		\begin{aligned}
			\partial_t \varrho_\mathrm{ex}=& -i \Big[ \frac{p^2}{2} + \frac{\gamma_q(t)}{2} q^2, \varrho_\mathrm{ex}\Big]  - D_p(t) [q,[q,\hat{\varrho}_\mathrm{ex}]] - \frac{i}{2} \gamma_p(t) [\hat{q},\{\hat{p},\hat{\varrho}_\mathrm{ex}\}]+D_q(t) [\hat{q},[\hat{p},\hat{\varrho}_\mathrm{ex}]].
		\end{aligned}
		\label{eq:appHuPazZhang}
	\end{align}
\end{widetext}
Above $\gamma_q(t)$, $\gamma_p(t)$, $D_p(t)$ and $D_q(t)$ are time-dependent parameters that depend on the bath properties and details of the coupling. 
It is a special property of the damped harmonic oscillator that all non-Markovian features are fully encoded in the time-dependence of the parameters without the need for non-local integral kernels in the master equation \cite{Paz94,JPiilo2004}. It is easy to check that \cref{eq:appHuPazZhang} is hermiticity and trace preserving. However, it should be noted that it is not a Lindblad master equation even in the Markovian limit for which the time-dependent parameters are replaced by their asymptotic values. The necessity for completely positive maps is strongly debated in the community and any physical quantum map should only guarantee the weaker condition of positivity \cite{sudarshan2005}. This is indeed the case for the above master equation (\ref{eq:appHuPazZhang}). 
\begin{figure}[b]
	\includegraphics{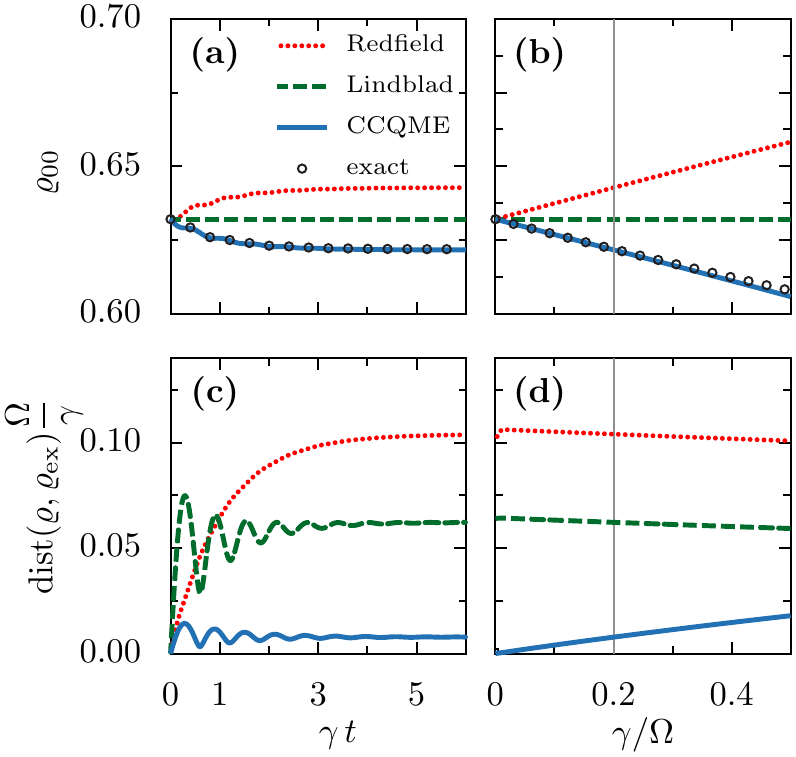}
	\caption{Analogue to Fig.~1 in the main text except for a higher bath temperature $T/\Omega=1$.} 
	\label{fig:app_main}
\end{figure}

The derivation of the above exact master equation and the details of the parameters are discussed in Ref.~\cite{KarrleinGrabert97}. Essentially there are two building blocks for the parameters. On the one hand there is the system's autocorrelation $G(t) = i \expval{[q(t),q(0)]}$ that satisfies the damped oscillator equation 
\begin{align}
	\ddot G(t) + \int_0^t \kappa(t-\tau) \dot{G}(\tau )\mathrm d\tau +  \Omega^2 G(t) = 0
\end{align}
with initial conditions $G(0)=0$ and $\dot G(0)=1$. The damping kernel $\kappa(\tau) = \sum_k (c_k^2/m_k \omega_k^2) \cos(\omega_k \tau)=\int_0^\infty (J(\omega)/\omega) \cos(\omega t)\mathrm d \omega$ is determined by the bath parameters. The expression in terms of the spectral density $J(\omega)$ holds in the continuum limit of the bath. In our work we consider an Ohmic spectral density with Drude cutoff 
\begin{align}
	J(\omega) =  \frac{\gamma\omega}{1 + (\omega/\omega_\mathrm{D})^2} ,
\end{align}
for which one finds an analytic solution of the autocorrelation in Laplace space \cite{KarrleinGrabert97},
\begin{align}
	\begin{split}
		\tilde G(z) &= \frac{1}{z^2 + z \tilde{\kappa}(z) + \Omega^2} \\
		& = \frac{\omega_\mathrm{D} + z}{z^3 + z^2 \omega_\mathrm{D} + z\ (\Omega^2 + \gamma \omega_\mathrm{D}) + \omega_\mathrm{D} \Omega^2},
		\label{eq:app_retGreenfc}
	\end{split}
\end{align} 
where the second line follows explicitly for the Drude bath.
By inverse Laplace transformation two of the four parameters are readily obtained,
\begin{align}
	\gamma_q(t) &= \frac{\ddot{G}(t)^2 - \dot{G}(t) \dddot{G}(t)}{\dot{G}(t)^2 - G(t) \ddot{G}(t)}, \\ 
	\label{eq:gammaq}
	\gamma_p(t) &= \frac{G(t) \dddot{G}(t) - \dot{G}(t) \ddot{G}(t)}{\dot{G}(t)^2 - G(t) \ddot{G}(t)},      
\end{align}
and are particularly temperature independent. For notational simplicity we use the dots above the $G$ to indicate the order of the time derivative.

On the other hand there is the bath's autocorrelation function $C(t) = \expval{B(t)B(0)}$ for the bath operator $B= \sum_k^\infty -c_k q_k$ that is well known from the Redfield formalism \cite{JThingaPHaenggi2012}. The diffusive parameters are given by 
\begin{align}
	D_q(t) &= \frac{1}{2} \ddot{K}_q(t) - K_p(t) + \gamma_q(t) K_q(t) + \frac{\gamma_p(t)}{2} \dot{K}_q(t), \\
	D_p(t) &= \frac{1}{2} \dot{K}_p(t) + \frac{\gamma_q(t)}{2} \dot{K}_q(t) + \gamma_p(t) K_p(t).   
	\label{eq:Dp}
\end{align}
Here $K_q(t)$ and $K_p(t)$ are real time influence kernels \cite{KarrleinGrabert97,HaakeReibold85} given by,
\begin{align}
	K_q(t) &= \int_0^t \mathrm{d}s  \int_0^t \mathrm{d}u\ \mathrm{Re}[C(|s- u|)]\ G(s)\ G(u), 	\label{eq:real-time-influence-kernel_Q} \\
	K_p(t) &= \int_0^t \mathrm{d}s  \int_0^t \mathrm{d}u\ \mathrm{Re}[C(|s- u|)]\ \dot G(s)\ \dot G(u). 	\label{eq:real-time-influence-kernel_P}
\end{align}

In this work we use the exact result to benchmark the canonically consistent quantum master equation (CCQME) derived in the main text. For the numerics we solve the master equations in the Markovian limit that is for times where the parameters $\gamma_q(t)$, $\gamma_p(t)$, $D_q(t)$, $D_p(t)$ and the Redfield rates/Lamb shifts are relaxed to their asymptotic values. 

\begin{figure}[t]
	\includegraphics{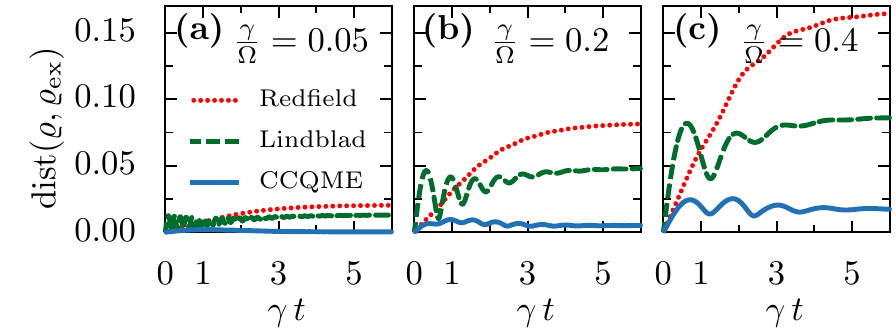}
	\caption{Error for Redfield (dotted red), Lindblad (dashed green), and CCQME (solid blue) calculated as trace distance to exact result as a function of scaled time for different values of the coupling strength (a)~$\gamma/\Omega=0.05$, (b)~$\gamma/\Omega=0.2$, and (c)~$\gamma/\Omega=0.4$. The initial state is $\varrho(0) =  \exp[-H_\mathrm{S}/\Omega]/Z_\mathrm{S}$ and the bath temperature is $T/\Omega=0.3$. For the simulation we take the Markovian master equations with Drude cutoff $\omega_\mathrm{D}/\Omega=5$. }
	\label{fig:app_error_coupling}
\end{figure}

Similar to the analysis in the main text we calculate the ground state population in \cref{fig:app_main}(a), (b) and in \cref{fig:app_main}(c), (d) the trace distance to the exact result,
\begin{align}
	\mathrm{dist}(\varrho, \varrho_\mathrm{ex}) = \frac{1}{2} \mathrm{Tr}(\sqrt{ (\varrho(t) - \varrho_\mathrm{ex}(t))^2}).
\end{align}
Here we consider a larger bath temperature $T/\Omega=1$ for which quantum effects are less relevant. Compared to the system time scale given by $1/\Omega$, in this regime the bath correlation function decays much faster and the Markov approximation is well justified. In this high-temperature regime the Redfield and Lindblad equations are usually expected to be valid approximations. This is shown in \cref{fig:app_main} where the results are close together and in particular the Redfield solution remains positive. However, strictly speaking also here both the Redfield and Lindblad result only remain valid for ultraweak coupling, whereas the CCQME accurately captures the ground state population also for finite coupling in \cref{fig:app_main} and overall has the smallest trace distance to the exact result [\cref{fig:app_main,fig:app_error_coupling}]. Contrastingly the Lindblad steady state predictions are independent of the coupling strength and the Redfield result increases with coupling strength as seen in \cref{fig:app_main}(b) and Fig. 1(b) [main text]. As described in the main text, for stronger coupling the effective temperature should increase and the higher lying states should get populated, a property only captured by the CCQME. 

In Fig.~\ref{fig:app_error_coupling} we plot the trace distance for different values of the system-bath coupling strength $\gamma$. By construction the steady state of the CCQME accurately captures finite coupling effects but it is not obvious that it also holds for the dynamics. In the main text we analyze the time averaged trace distance as a function of bath temperature and coupling strength. Here we give an overview for the dynamics of the trace distance for different values of the coupling strength. All panels show the same results qualitatively, with the CCQME outperforming the other QMEs for the error. Moreover, for the CCQME the trace distance decreases monotonically with smaller values of the coupling not only in the steady-state regime but also throughout the evolution. \\

\subsection{CCQME regime of validity}
\begin{figure}[t]
	\includegraphics{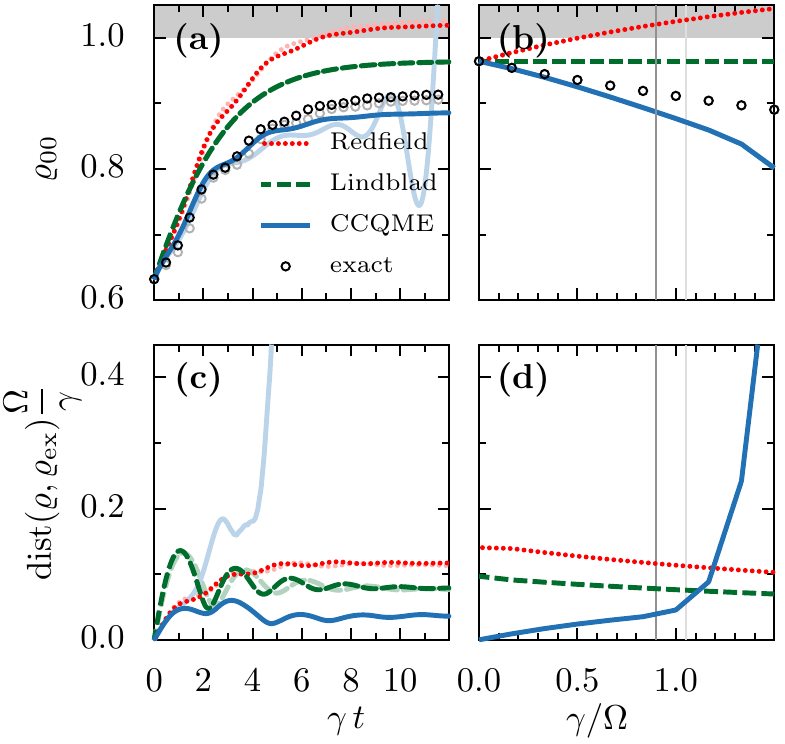}
	\caption{Analogue to Fig. 1 of the main text except for $\gamma/\Omega=0.9$ (dark lines and symbols), $\gamma/\Omega=1.01$ (light lines and symbols), and $\omega_\mathrm{D}/\Omega=1$. }
	\label{fig:app_validity}
\end{figure}

For weak but finite coupling, i.e.~$\gamma/\Omega\leq 0.5$ the CCQME very accurately agrees with the exact dynamics (see main text). That is because the CCQME is a second-order consistent master equation whose coupling strength $\gamma$ is a perturbative (control) parameter, that governs the precision of the simulated dynamics. If we increase $\gamma$ beyond the second order coupling regime, i.e.~$\gamma/\Omega=0.9$, the CCQME steady state starts to deviate significantly from the exact result but the solution is still positive 
[see dark lines and symbols in \cref{fig:app_validity}]. For strong coupling $\gamma/\Omega>1$ the approximations made for the CCQME are no longer valid and the solution may violate positivity [see light lines and symbols in \cref{fig:app_validity}].

\subsection{CCQME for a system connected to multiple baths}
\begin{figure}[b]
	\includegraphics{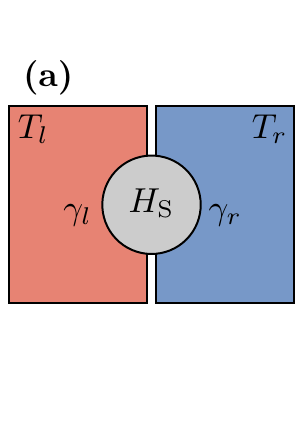}
	\includegraphics{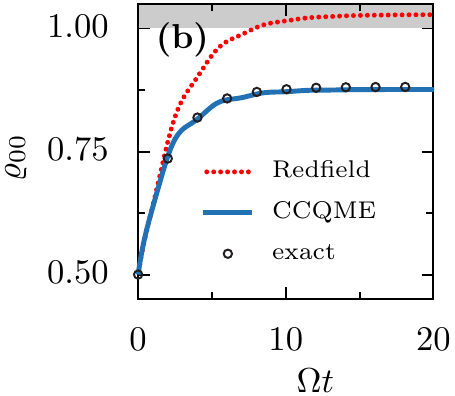}
	\caption{Harmonic oscillator coupled to two independent baths illustrated in panel (a). Dynamics for the ground state population when the system is initiated in a superposition of the ground and first excited state $\ket{\psi(0)} = (1/\sqrt{2})\, (\ket 0 + \ket 1)$ under the Redfield equation (dotted red), boundary driven CCQME (solid blue), and the exact solution (open black circles) in panel (b). For the simulation $N=20$ first levels are taken. The bath parameters are $\gamma_l/\Omega=\gamma_r/\Omega = 0.2$, $T_l/\Omega=0.3$, $T_r/\Omega=0.5$, and $\omega_\mathrm{D}/\Omega=5$.} 
	\label{fig:app_noneq}
\end{figure}

In this section we implement the CCQME to a minimal transport setup wherein the harmonic oscillator system is driven by two independent thermal baths as depicted schematically in \cref{fig:app_noneq}(a). For uncorrelated baths the damping kernel and bath correlation functions simply add up such that the autocorrelation function \cref{eq:app_retGreenfc} and the real time influence kernels \cref{eq:real-time-influence-kernel_Q,eq:real-time-influence-kernel_P} are altered to
\begin{align}
	\tilde G(z) &= \frac{1}{z^2 + z \sum_i \tilde \kappa^i(z) + \Omega^2}, \\
	K_q(t) &= \int_0^t \mathrm{d}s  \int_0^t \mathrm{d}u\ \sum_i \mathrm{Re}[C^i(|s- u|) ]\ G(s)\ G(u), \\
	K_p(t) &= \int_0^t \mathrm{d}s  \int_0^t \mathrm{d}u\ \sum_i \mathrm{Re}[C^i(|s- u|)]\ \dot G(s)\ \dot G(u). 	
\end{align}
The remaining structure of equations~\eqref{eq:gammaq}-~\eqref{eq:Dp} remain the same allowing us to simulate the exact quantum master equation~\cref{eq:appHuPazZhang} for the nonequilibrium scenario.

Following the line of arguments of the main text we start with the Redfield equation
\begin{align}
	\partial_t \varrho(t) = -i [H_\mathrm{S},\varrho(t)] +  (\mathcal{R}_{\infty}^{l} +  \mathcal{R}_{\infty}^{r})[\Lambda_t^0[\varrho(0)]],
\end{align}
where the Redfield operators for the left and right bath respectively act on the initial state. Replacing the initial state according to the inverse map,
\begin{align}
	\Lambda_t^0[\varrho(0)] \simeq ( \mathbb{I} - \bar Q^{l} - \bar Q^{r} )[\varrho(t)], 
\end{align}
wherein the superoperators are obtained from~\cref{eq:app_formal_deriv,eq:app_Q2_redfield} we arrive at the boundary driven CCQME,
\begin{align}
	\begin{aligned}
		\partial_t \varrho(t) =&-i \left[H_\mathrm{S}, \varrho(t)\right] + (\mathcal{R}_{\infty}^{l} +  \mathcal{R}_{\infty}^{r}) [ (\mathbb{I} - \bar Q^{l} - \bar Q^{r} )[\varrho(t)] ] .
	\end{aligned}
	\label{eq:app_noneq_CCQME}
\end{align}
It's important to note here that even though the left ($l$) and right ($r$) baths are uncorrelated the dissipators beyond weak coupling (second order) are not. This property can also be observed in the exact formulation or fourth-order QMEs~\cite{CaoJCP,JThingnaWJSheng2014} and is simple to note in our boundary driven CCQME.

For Ohmic baths with a Drude cutoff the dynamics for the ground state population is depicted in \cref{fig:app_noneq}(b). As expected for low temperatures and finite coupling the Redfield theory shows an unphysical behaviour with the ground state population exceeding one. On the other hand, the CCQME shows a surprisingly good agreement with the exact solution demonstrating that our approach is easily extendable to study the dynamics of boundary driven quantum systems. In future works we will connect it with recent advances in evaluating the asymptotic state of nonequilibrium quantum systems driven by multiple reservoirs~\cite{ThingnaPRE13} or an external drive~\cite{TShirai2016, CaoPRL19}.

\subsection{Spin-boson model}

As compared to other numerically exact approaches such as the hierarchy equation of motion (HEOM) approach \cite{HEOM89, FruchtmanLambertGauger2016}, the CCQME does not require more resources as compared to the Redfield formalism. Therefore it can be applied to larger systems and for low bath temperatures at which the resources for the HEOM approach scale exponentially.

\begin{figure}[b!]
	\includegraphics{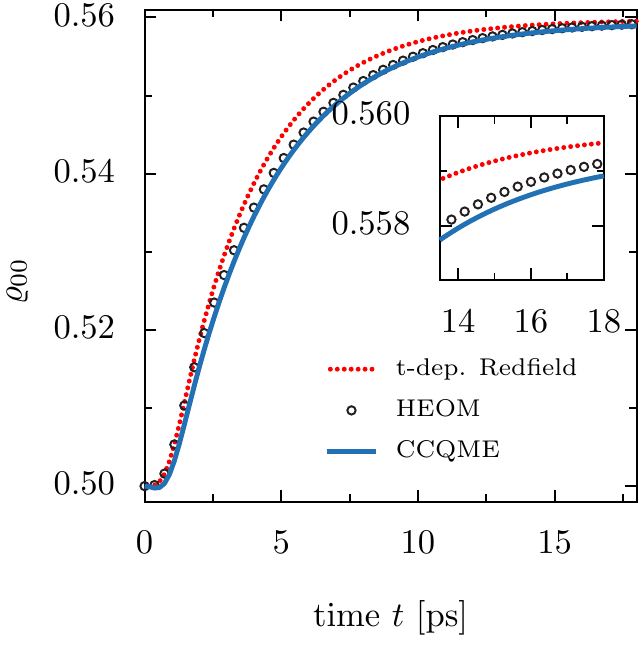}
	\caption{Comparison of the spin-boson ground-state dynamics with the HEOM data taken from \cite{FruchtmanLambertGauger2016} (open circles), time-dependent Redfield (dotted red), and the CCQME result (solid blue). The parameters are chosen as $\varepsilon=\pi/2\, \mathrm{ps}^{-1}$, $T=50\, K$, $\lambda=0.01485\, \mathrm{ps}^{-1}$, and $\omega_c=2.2\, \mathrm{ps}^{-1}$.}
	\label{fig:app_spin-boson}
\end{figure}

In this section we compare the CCQME with the HEOM for the spin-boson model. The total system--bath Hamiltonian reads as
\begin{align}
	H_\mathrm{tot} = \frac{\varepsilon}{2} \sigma_x + \sigma_z \sum_k g_k (a^\dagger_k + a_k) + \sum_k \omega_k a^\dagger_k a_k
\end{align}
and we consider an Ohmic spectral density with exponential cutoff 
\begin{align}
	J(\omega) = \sum_k g^2_k \delta(\omega - \omega_k) = \lambda (\omega/\omega_c) e^{-\omega^2/\omega_c^2}.
\end{align}

For the Redfield dynamics we take the time dependent rates $W(E, t) =\int_0^t C(\tau) e^{-i E \tau} \mathrm d \tau$ with bath correlation function given by
\begin{align}
	C(\tau) = \int_0^\infty J(\omega) \big[ \coth(\beta\omega/2)\cos(\omega \tau) -i \sin(\omega \tau) \big] \mathrm d\omega.
\end{align}
Here for ultraweak coupling and high bath temperature, the Redfield equation guarantees positivity as shown explicitly for the ground state population in \cref{fig:app_spin-boson}. However it does not match the numerically exact HEOM dynamics. For the CCQME we combine the time dependent Redfield equation with the static $Q$-tensor (see main text) and obtain accurate dynamics without the computational complexity of HEOM.

\subsection{Ising chain}

\begin{figure}
	\includegraphics{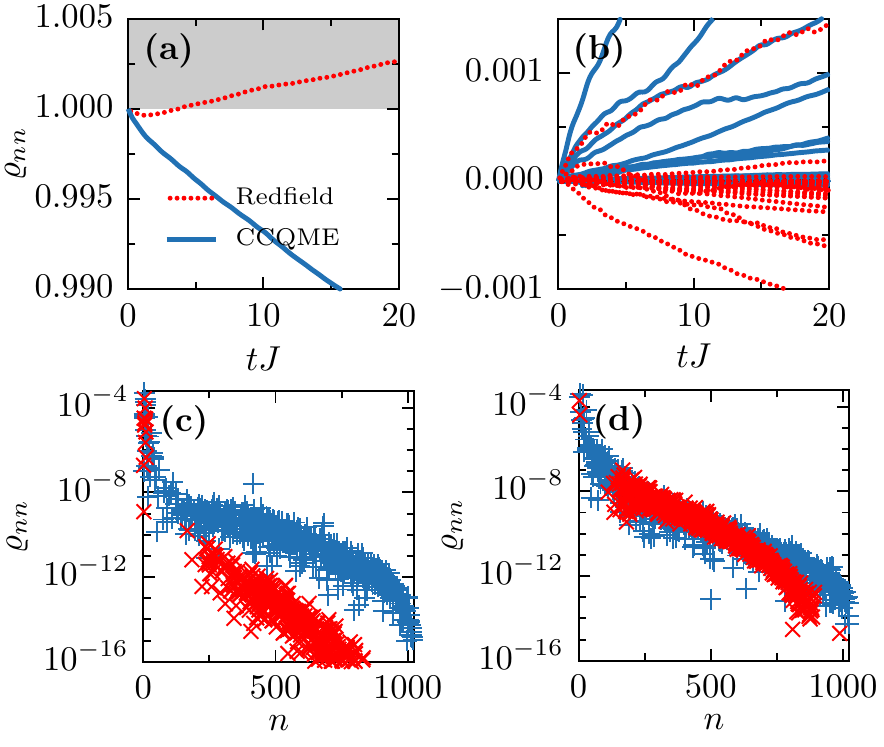}
	\caption{Many-body populations of a spin chain of length $L=10$ coupled locally to a Lorentz-Drude bath for (a)~the many-body ground state and (b) all excited states as a function of time $t$.  Snapshot of the populations at (c) $t=1/J$ and (d) $t=20/J$, red (blue) crosses depict Redfield (CCQME). On the logarithmic scale, only positive populations can be shown. Redfield eventually predicts negative populations that are of the order of the coupling strength, while the CCQME remains mostly positive at long times. The  parameters are chosen as $\gamma=0.04J$, $T=0.9J$, and $\omega_\mathrm{D}=5J$. 
	}
	\label{fig:app_occups_spinchain}
\end{figure}

Having benchmarked the performance for a single-spin-boson model, here we demonstrate the performance of the CCQME for a genuine many-body system, where we choose
an Ising chain of length $L$ with open boundary conditions and external fields in $x$- and $z$-directions coupled to a bosonic Lorentz-Drude bath. The corresponding total system-bath
Hamiltonian reads,
\begin{align}
	\begin{split}
		H_\mathrm{tot} =&-J\sum_{i=1}^{L-1} \sigma_z^{i}\sigma_z^{i+1} +  \sum_{i=1}^{L}\left( h_x^{i} \sigma_x^{i} + h_z^{i} \sigma_z^{i}\right) \\
		&+ \sigma_z^{L/2} \sum_k g_k (a^\dagger_k + a_k) + \sum_k \omega_k a^\dagger_k a_k,
	\end{split}
\end{align}
where the bath is coupled locally via a dephasing coupling to the central site. In order to avoid near degeneracies in the 
many-body spectrum we choose slowly twisting fields according to
\begin{align}
	h_x^{i} =& 0.8J+\frac{i-1}{L-1} 0.2 J,\\
	h_z^{i} =& 0.7J-\frac{i-1}{L-1} 0.2 J,
\end{align}
that breaks the underlying symmetries lifting degeneracies.

We consider a Lorentz-Drude spectral density
\begin{align}
	J(\omega) = \sum_k g^2_k \delta(\omega - \omega_k) = \frac{\gamma\omega}{1 + (\omega/\omega_\mathrm{D})^2},
\end{align}
with coupling strength $\gamma$ and cut-off $\omega_\mathrm{D}$.
Such a spectral density is not easy to treat via HEOM \cite{FruchtmanLambertGauger2016} or pseudomode mapping \cite{Imamoglu94,Garraway97,Mazzola09,Strunz20} (also known as mesoscopic leads approach). 
Hence the relatively large system size of $L=10$ (and Hilbert space dimension $2^{10}=1024$) is unaccessible for those methods using a machine with $64\mathrm{GB}$ RAM that we use, while we are still able to solve the CCQME.

In Fig.~\ref{fig:app_occups_spinchain}, we plot the dynamics of the populations $\varrho_{nn}$ of the many-body eigenstates for a chain of length  $L=10$ both for the Redfield master equation as well as for the CCQME
, starting in the pure many-body ground state $\varrho(0)=\ketbra{\psi_0}{\psi_0}$. Remarkably, even though the Redfield solution is unphysical, with many negative populations of the excited states [Fig.~\ref{fig:app_occups_spinchain}(b)] and a ground-state po\-pu\-lation above unity [Fig.~\ref{fig:app_occups_spinchain}(a)], the ground-state po\-pu\-lation for the CCQME solution is physical at all times, and positivity violation of the excited state populations is rare and of small magnitude. At $t=20/J$ the negative populations in Redfield sum up to $\sum_{n, \varrho_{nn}<0} \varrho_{nn}(t) \approx -4.3\cdot10^{-2}$ which is of the order of the coupling strength $\gamma/J=4\cdot10^{-2}$. For the CCQME the negative populations only arise on second order in the coupling strength $\sum_{n, \varrho_{nn}<0} \varrho_{nn} \approx- 6.7\cdot10^{-5}$ which is an advantage by three orders of magnitude. The long-time state of the CCQME in Fig.~\ref{fig:app_occups_spinchain}(d) is also consistent with ETH with populations being a smooth function of energy and small fluctuations around it.

\end{document}